\documentclass[twocolumn,preprintnumbers,superscriptaddress,amsmath,showpacs,amssymb]{revtex4}
\usepackage{epsfig,graphicx,times}
\usepackage{amstext}
\usepackage{amsmath}            
\usepackage{amssymb}            
\usepackage{graphicx}           
\usepackage{latexsym}
\usepackage{bm}

\begin{document}

\title{Tomographic measurements on superconducting qubit states}
\author{ Yu-xi Liu}
\affiliation{Frontier Research System,  The Institute of Physical
and Chemical Research (RIKEN), Wako-shi 351-0198, Japan}
\author{L.F. Wei}
\affiliation{Frontier Research System,  The Institute of Physical
and Chemical Research (RIKEN), Wako-shi 351-0198, Japan}
\affiliation{Institute of Quantum Optics and Quantum Information,
Department of Physics, Shanghai Jiaotong University, Shanghai
200030, P.R. China }
\author{Franco Nori}
\affiliation{Frontier Research System,  The Institute of Physical
and Chemical Research (RIKEN), Wako-shi 351-0198, Japan}
\affiliation{Center for Theoretical Physics, Physics Department,
Center for the Study of Complex Systems, The University of
Michigan, Ann Arbor, Michigan 48109-1120, USA}

\date{\today}

\begin{abstract}
We propose an approach to reconstruct any superconducting charge
qubit state by using quantum {\it state} tomography. This
procedure requires a series of measurements on a large enough
number of identically prepared copies of the quantum system.  The
experimental feasibility of this procedure is explained and the
time scales for different quantum operations are estimated
according to experimentally accessible parameters. Based on the
{\it state} tomography, we also investigate the possibility of the
{\it process} tomography.

\end{abstract}

\maketitle \pagenumbering{arabic}
\section{introduction}
 The generation of superpositions of  macroscopic quantum
states in superconducting devices~\cite{jr,Nakamura,pashkin,
chiorescu,han} have motivated further research on quantum
information processing in these systems. Two types of
superconducting qubits based on Josephson junction devices have
been proposed and experimentally demonstrated. One involves  two
Cooper-pair charge states in a small superconducting island
connected to a circuit by a Josephson tunnel junction and a gate
capacitor (see, e.g.,~\cite{Nakamura,pashkin,makhlin}). An
alternative approach is based on the phase states of a Josephson
junction or the flux states in a ring superconducting
structure~\cite{ioffe,chiorescu,han}. Further, experimental
observations on quantum oscillations and the demonstration of
conditional gate operations in two coupled charge
qubits~\cite{pashkin} are necessary first steps towards future
realizations of quantum information processors.

A crucial step in quantum information  processing is the
measurement of the output quantum states. However, a quantum state
cannot be ascertained by a single quantum measurement. This is
because quantum states may comprise many complementary features
which cannot be measured simultaneously and precisely due to
uncertainty relations. However, all complementary aspects can in
principle be observed  by a series of measurements on a large
enough number of identically prepared copies of the quantum
system. Then we can reconstruct a quantum state from such a
complete set of measurements of system observables (i.e., the
quorum~\cite{qu}). Such a procedure is called ``reconstruction of
quantum states" or Quantum State Tomography (QST).

Quantum state tomography is not only important for quantum
computation, which requires the verification of the accuracy of
quantum operations, but it is also important for fundamental
physics. Many theoretical studies for tomographic reconstruction
of quantum states have been done, e.g.
references~\cite{theor,dfv,long,Ariano}. Experimentally,
tomography has been investigated for a variety of systems,
including: e.g., the vibrational state of molecules~\cite{dunn},
the motional quantum state of a trapped atom~\cite{lei,kur},
two-photon states~\cite{white}, the electromagnetic
field~\cite{smithey}, and rare-earth-metal-ion-based solid-state
qubit~\cite{jjl}, the two-qubit states in the trapped
ions~\cite{cfr}. The quantum states of multiple spin-$\frac{1}{2}$
nuclei have also been measured in the high-temperature regime
using NMR techniques~\cite{chuang,miquel,cory}.

For continuous variable cases (e.g., the molecular vibrational
mode~\cite{dunn}, motional quantum states of a trapped
ion~\cite{lei,kur}, a single-mode~\cite{smithey} of the
electromagnetic field), the quantum states can be known by the
tomographic measurement of their Wigner function. For the discrete
variable case (e.g., in NMR systems), the measurements on the
density matrix in NMR experiments are realized by the NMR spectrum
of the linear combinations of ``product operators", i.e. products
of the usual angular momentum operators~\cite{cory}.

Based on the state tomography, a quantum ``black box" connected to
an unknown external reservoir can also be characterized. This
``black box" transfers any known input state to an unknown output
state. The determination of the quantum transfer function for this
`black box" is called quantum process tomography~\cite{il}. This
procedure needs to input a large enough number of different known
states into the ``black box", then to make tomographic
measurements on output states, finally to obtain the quantum
transfer function, which determines the ``black box". This
procedure would be very important for the case when the noisy
channel is unclear. Process tomography has been experimentally
realized, e.g., in  optical systems~\cite{mwm}, NMR~\cite{ysw}.

To our knowledge, there is no adequate theoretical analysis or
experimental demonstration for the reconstruction of qubit states
in solid state systems, besides our recent work in
Ref.~\cite{liu}. There, we considered a very general class of spin
Hamiltonians used to model generic solid state systems~\cite{liu}.
Here, the emphasis is not on a general model but on a specific
system: superconducting qubits. Recent technical progress makes it
possible to realize quantum control in superconducting quantum
devices and ascertain either the charge~\cite{Nakamura,pashkin} or
the flux~\cite{chiorescu} qubit states. Furthermore, practical
experiments on quantum computing require the knowledge of the full
information of the quantum state, so the reconstruction of quantum
states in solid state systems is a very important issue.

In this paper, we analyze how to reconstruct charge qubit states
in superconducting circuits. In principle, if all qubits can be
measured at same time in superconducting circuits as in optical
systems (e.g., Ref.~\cite{dfv}), then only single-qubit operations
are enough to assist the implementation of the reconstruction of
any multiple-qubit state. However, our proposal only considers one
qubit measurement at a time. This is because simultaneous
measurements of many qubits are currently very difficult to
implement~\cite{comments} in superconducting circuits. Another
reason is that simultaneous measurements require many probes in
contact with qubits, inducing more noisy channels. These multiple
noisy channels will quickly reduce the coherence of the qubit
states, decreasing the accuracy of the reconstructed states. So
for two-qubit and multi-qubit state tomography, appropriate
two-qubit operations are necessary, due to the constraint of a
single-qubit measurement at a time.

Although our analysis of the tomographic reconstruction of charge
qubit states might seem somewhat similar to the one used for NMR
systems~\cite{chuang,miquel,cory}, there are significant
differences on how to realize the state tomography in Josephson
junction (JJ) charge qubits. For example, NMR QST (like optical
QST) also only involves single-qubit operations. A question we
will focus on is the following: is it possible to do QST with the
currently accessible experimental capability on JJ qubits? In view
of the short relaxation and decoherence times, it is also
necessary to estimate quantum operation times required for
reconstructing charge qubit states. In particular, it is not
trivial to find an appropriate two-qubit operation to realize all
two and multiple qubit measurements.

Here, we theoretically analyze in detail the necessary
experimental steps for the tomographic reconstruction of
dc-SQUID-controlled charge qubit states. This analysis can be
easily generalized to other proposals of controllable
superconducting qubits (e.g., flux and phase qubits).  In Sec. II,
the reconstruction of single-qubit states is described in detail.
The time scales of operations for measurements of all three
unknown matrix elements are also estimated by using currently
accessible experimental parameters.  In Sec.~III, all operations
required to reconstruct two-qubit states are given, the time
scales for the first and second qubit measurements are estimated
using experimentally accessible parameters. In Sec.~IV, using an
example, we generalize our two-qubit tomography to the
multiple-qubit case. Finally in Sec.~V, we discuss the ``process
tomography" of singe-qubit charge systems based the ``state
tomography".  Sections III, IV, and V contain our most important
results. The conclusions and further discussions are given in
sections VI and VII, respectively.

\section{reconstruction of single-qubit states}

The content of this section on single-qubit operations and the
reconstruction of single-qubit states is known~\cite{book} to
specialists in the optical, NMR and other areas
(e.g,~\cite{qu,theor,dfv,long,Ariano,dunn,lei,kur,white,smithey,jjl,cfr,chuang,miquel,cory}),
where the QST is extensively studied. But here we specify a
detailed description of the steps needed for the experimental
realization of the tomographic reconstruction for charge qubit
states. This should be helpful to solid state experimentalists who
are not specialists on the QST.

\subsection{Theoretical model and single-qubit states}
We consider a controllable dc-SQUID system which consists of a
small superconducting island with $n$ excess Cooper-pair charges,
connected by two nominally-identical ultra-small Josephson
junctions; each having capacitance $C^{0}_{J}$ and Josephson
coupling energy $E^{0}_{J}$. A control gate voltage $V_{g}$ is
coupled to the Cooper-pair island by a gate capacitance $C_{g}$.
The qubit is assumed to work in the charge regime, e.g., the
single-electron charging energy $E_{C}=e^2/2(C_{g}+2C^{0}_{J})$
and Josephson coupling energy $E^{0}_{J}$ satisfy the condition
$E_{C}\gg E_{J}$. If the applied gate voltage range $V_{g}$ is
near a value $V_{g}=e/C_{g}$, only two charge states, denoted by
$n=0$ and $n=1$, play a key role, then this charged box is reduced
to a two-level system (qubit) whose dynamical evolution is
governed by the Hamiltonian~\cite{makhlin,y}
\begin{equation}\label{eq:1}
H=-\,\frac{1}{2}\,\delta\!E_{\rm ch}(n_{\rm
g})\,\sigma_{z}-\frac{1}{2}E_{\rm J}(\Phi_{x})\,\sigma_{x},
\end{equation}
where we adopt  the convention of  charge states
$|0\rangle=|\!\uparrow\rangle$ and
$|1\rangle=|\!\downarrow\rangle$. The charge energy $\delta\!
E_{\rm ch}(n_{\rm g})=4\,E_{\rm C}(1-2n_{\rm g})$ with $n_{\rm
g}=C_{\rm g}V_{\rm g}/2e$ can be controlled by the gate voltage
$V_{\rm g}$. The Josephson coupling energy $E_{\rm
J}(\Phi_{x})=2\,E^{0}_{\rm J}\cos(\pi\Phi_{x}/\Phi_{0})$ is
adjustable by the external flux $\Phi_{x}$, and $\Phi_{0}=h/2e$ is
the flux quantum. Our goal here is to determine any single charge
qubit state by the controllable dynamical operation governed by
the Hamiltonian~(\ref{eq:1}).

Any single-qubit state (mixed or pure) can  be represented by a
density matrix operator in a basis
$\{|0\rangle=|\!\uparrow\rangle,\,|1\rangle=|\!\downarrow\rangle
\}$ as
\begin{subequations}\label{eq:2}
\begin{eqnarray}
\rho&=&\left(\begin{array}{cc}\rho_{00} &\rho_{01}\\
\rho_{10}&\rho_{11} \end{array}\right)
=\frac{1}{2}\sum_{k=0,x,y,z}r_{k}\,\sigma_{k},
\end{eqnarray}
 or
\begin{eqnarray}
\rho&=&\rho_{00}\,|0\rangle\langle
0|+\rho_{01}\,|0\rangle\langle
1|+\rho_{10}\,|1\rangle\langle 0|\nonumber\\
&+& \rho_{11}\,|1\rangle\langle 1|,
\end{eqnarray}
\end{subequations}
where $\sigma_{k=x,y,z}$ are Pauli operators and $\sigma_{k=0}$ is
an identity operator. Four real parameters $r_{k}\,\, (k=0,x,y,z)$
can be expressed as
\begin{eqnarray*}
r_{0}&=& \rho_{00}+\rho_{11}, \,\, \,\,\,\, \,\,\,\,\,\,\,\,\,\,r_{x}= \rho_{01}+\rho_{10}, \nonumber\\
r_{y}& =&i (\rho_{01}-\rho_{10}),\,\,\,\,\,\, \,\,\,\,\,r_{z}=
\rho_{00}-\rho_{11}.
\end{eqnarray*}
The normalization condition $\rho_{00}+\rho_{11}=1$ ensures that
the qubit~(\ref{eq:2}) can actually be determined by three real
parameters $ r_{x}$, $r_{y}$, $r_{z}$ corresponding~\cite{book} to
a Bloch vector $\overrightarrow{r}$, which satisfies the condition
$|\overrightarrow{r}|\leq 1$ (see Fig.\ref{fig1}(a)). The state
$\rho$ is pure if and only if $|\overrightarrow{r}|=1$. When the
state $\rho$ is pure, the Bloch vector $\overrightarrow{r}$
defines a point on the unit three-dimensional sphere.

These three coefficients $r_{k} \,\,(k=x,\,y,\,z)$  can be
obtained from measurements of $\sigma_{x}$, $\sigma_{y}$,
$\sigma_{z}$. The correspondence between these three measurements
and the coefficients $r_{k}$ is given by
\begin{equation*}
r_{k}\,=\,{\rm Tr}(\rho\,\, \sigma_{k}),
\end{equation*}
due to the relation ${\rm Tr}(\sigma_{i}\sigma_{j})=2\delta_{ij}$,
where $\delta_{ij}$ is the Kronecker delta.

\subsection{Quantum operations and measurements on single-qubit states}

In principle, the state of the charge qubit can be read  by a
single-electron transistor (SET)~\cite{Nakamura,pashkin,y} coupled
capacitively to a charge qubit. Here we consider the ideal case in
which the SET is coupled to the qubit only during the measurement.
When the SET is coupled to the qubit, the dissipative current $I$
flowing through the SET is proportional to the probability of a
projective operator measurement $|1\rangle\langle 1|$ on the qubit
state, which has actually been applied by the
experiment~\cite{Nakamura,pashkin}. The $|1\rangle\langle 1|$
measurement is equivalent to a $\sigma_{z}$ measurement on the
state $\rho$,
\begin{equation*}
p_{1}\,=\,{\rm Tr}(\rho \,|1\rangle\langle
1|)\,=\,\frac{1}{2}[1-{\rm Tr}(\rho\,\sigma_{z})]\,=\,\rho_{11}
\end{equation*}
due to the relation
\begin{equation*}
|1\rangle\langle 1|\, =\,\frac{1}{2}(\sigma_{0}-\sigma_{z}).
\end{equation*}
 The parameters
$r_{0}$ and $r_{z}$ can be determined by the result of the
measurement $|1\rangle\langle 1|$,  together with the
normalization condition.

We can also relate the two other measurement operators,
$\sigma_{x}$ and $\sigma_{y}$, to the operator $|1\rangle\langle
1|$ (essentially $\sigma_{z}$), which is the measurement
experimentally realized in the charge qubits. This is because the
current $I$ flowing through the SET is  sensitive to the charge
state $|1\rangle$, so the single qubit operations have to be
performed so that the desired parameter $r_{x}$ or $r_{y}$ is
transformed to the measured diagonal positions.

Now, we describe the steps to measure  $r_{x}$ or $r_{y}$. Let us
first choose the external flux $\Phi_{x}=0$ and suddenly drive the
qubit to the degeneracy point for a time
$$t_{x}=\frac{\hbar\pi}{2\,E_{\rm J}(0)}=\frac{\hbar\pi}{4\,E^{0}_{\rm J}}$$
such that the qubit state can be rotated $-\pi/2$ along the $x$
direction, here $E_{\rm J}(0)=E_{\rm J}(\Phi_{x}=0)$.

\begin{figure}
\includegraphics[bb=20 140 680 700, width=9.0 cm,
clip]{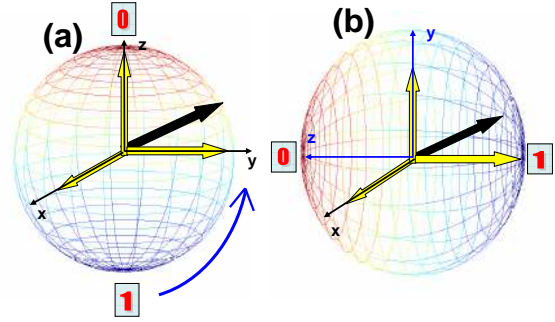} \vspace{-2.5cm} \caption[]{(Color on line) The
black Bloch vector indicates a qubit state; the (yellow arrows)
$r_{x},\, r_{y}\,$ and $r_{z}$ represent the three components of
the Bloch vector along the $x,\,y$ and $z$ axes. The $0$ and $1$
in the north and south poles of the Bloch sphere denote the
measured states $|0\rangle$ and $|1\rangle$, respectively. The
measurement instrument is attached to a pole (e.g., ``1") of the
sphere. A $-\pi/2$ rotation of the qubit state along the $x$
direction is equivalent to a $\pi/2$ rotation of the measuring
instrument along the $x$ direction.}\label{fig1}
\end{figure}

The probability $p_{2}$ of the measurement $|1\rangle\langle 1|$
on the rotated state is
\begin{eqnarray}\label{eq:p2}
p_{2}&=&{\rm
Tr}\left(R_{x}(t_{x})\,\,\rho\,\,R^{\dagger}_{x}(t_{x})\,|1\rangle\langle
1|\right)\nonumber\\
&=&{\rm
Tr}\left(\,\exp\left\{i\frac{\pi}{4}\sigma_{x}\right\}\,\rho\,
\exp\left\{-i\frac{\pi}{4}\sigma_{x}\right\}|1\rangle\langle
1|\,\right)\nonumber\\
&=& {\rm Tr}\left(\rho\,
\exp\left\{-i\frac{\pi}{4}\sigma_{x}\right\}|1\rangle\langle
1|\,\,\exp\left\{i\frac{\pi}{4}\sigma_{x}\right\}\,\right)\nonumber\\
&=&\frac{1}{2}\,(1+r_{y}),
\end{eqnarray}
where
$R_{x}(t_{x})=\exp\left\{iE_{J}(0)\sigma_{x}t_{x}/2\hbar\right\}$,
Eq.~(\ref{eq:p2}) means that the measurement $|1\rangle\langle 1|$
on the state rotated $-\pi/2$ along the $x$ direction is
equivalent to the measurement $\sigma_{y}$, and the rotation
$-\pi/2$ of the qubit is equivalent to  an inverse rotation of
the measuring instrument, see Fig. ~\ref{fig1}.

\begin{figure}
\includegraphics[bb=20 20 680 740, width=9.0 cm,
clip]{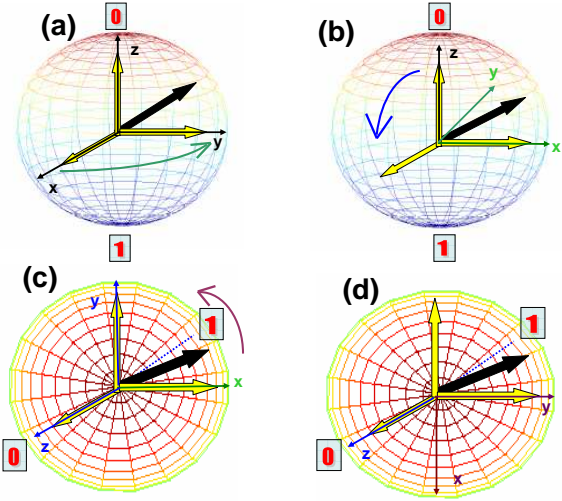} \vspace{-2.5cm} \caption[]{(Color on line) The
Bloch vector is the same as  Fig.~\ref{fig1}.  A $-\pi/2$ rotation
of the qubit along the $y$ direction is equivalently realized by
the rotation  $\pi/2$ of the measuring instrument along the $z$
direction (from (a) to (b)), then a $\pi/2$ rotation along the $x$
direction (from (b) to (c)), and a $3\pi/2$ rotation along $z$
direction (from (c) to (d)).}\label{fig1.2}
\end{figure}

In order to make the third  measurement $\sigma_{x}$,  the qubit
state needs now to be rotated $-\pi/2$ (or $\pi/2$) along the $y$
direction. This  can be done (e.g., $-\pi/2$ rotation) as follows:
\begin{description}

\item  (i) \,\,\,Set $\Phi_{x}=\Phi_{0}/2$ and $n_{\rm g}=0$; let
the system evolve a time $t_{z,1}=\hbar\pi/8E_{\rm C}$ such that a
rotation of $-\pi/2$ along the $z$ direction is realized.

\item  (ii)\,\, After the time $ t_{z,1}$, set $\Phi_{x}=0$ and
$n_{\rm g}=1/2$ and let the system evolve a time period
$t_{x,1}=\hbar\pi/2E_{\rm J}(0)=\hbar\pi/4E^{0}_{\rm J}$ such that
the system rotates $-\pi/2$ along the $x$ direction.

\item  (iii)\,\, Set $\Phi_{x}=0$ and $n_{\rm g}=1/2$ again and
let the system evolve a time $ t_{z,2}=3\hbar\pi/8E_{\rm C}$ and a
rotation $-3\pi/2$ along the $z$ direction is obtained.
\end{description}
Combining the above three steps, shown in Fig.~\ref{fig1.2}, a
$-\pi/2$ rotation of the qubit along the $y$ direction is
realized.
\begin{description}
\item  (iv)\,\, After the above rotations, a measurement
$|1\rangle\langle 1|$ on this rotated state must be made, which is
equivalent to measuring $\sigma_{x}$. Then, the measured
probability becomes
\begin{eqnarray*}
p_{3}&=&{\rm Tr}(\,R_{z,x,z}\,\rho\, R^{\dagger}_{z,x,z}\,|1\rangle\langle 1|\,)\nonumber\\
&=&{\rm Tr}\left(\,\exp\left\{i\frac{\pi}{4}\sigma_{y}\right\}\rho
\exp\left\{-i\frac{\pi}{4}\sigma_{y}\right\}|1\rangle\langle
1|\,\right)\nonumber\\
&=&\frac{1}{2}\,(1+r_{x})
\end{eqnarray*}
\end{description}
with $R_{z,x,z}=R_{z}(t_{z,1})R_{x}(t_{x,1})R_{z}(t_{z,2})$, and
\begin{eqnarray*}
R_{z}(t_{z,1})&=&\exp\left\{i \frac{2E_{\rm C}}{\hbar}\sigma_{z}t_{z,1}\right\}
=\exp\left\{i\frac{\pi}{4}\sigma_{z}\right\},\\
R_{x}(t_{x,1})&=&\exp\left\{i\frac{E^{0}_{\rm
J}}{\hbar}\sigma_{x}t_{x,1}\right\}=\exp\left\{i\frac{\pi}{4}\sigma_{x}\right\},\\
R_{z}(t_{z,2})&=&\exp\left\{i \frac{2E_{\rm
C}}{\hbar}\sigma_{z}t_{z,2}\right\}=\exp\left\{i\frac{3\pi}{4}\sigma_{z}\right\}.
\end{eqnarray*}
We explained how to measure the single qubit states by single
qubit operations and measuring $|1\rangle\langle 1|$. Below, we
give an example that shows a reconstructed single-qubit state can
be graphically represented, and we further give estimates of the
operation times to obtain each of the matrix elements of
single-qubit states.

\subsection{An example}
\begin{figure}
\includegraphics[width=7.0cm]{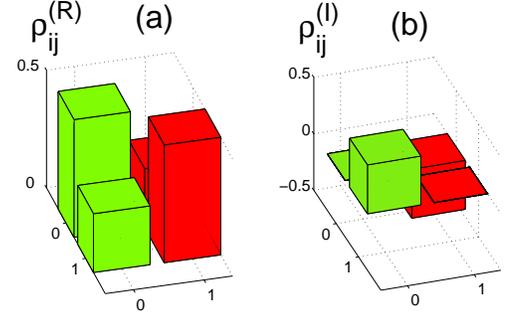}
\caption[1]{Graphical representation of the density matrix $\rho$
for single-qubit states, see the example explained in section II.
The real $\rho_{ij}^{\rm (R)}$ and imaginary $\rho_{ij}^{\rm (I)}$
parts of the density matrix elements $\rho_{ij}=\langle
i|\rho|j\rangle$ $(i,\, j=0,1)$ are plotted in (a) and (b),
respectively.}\label{fig3}
\end{figure}

The three measurement results ($p_{1},\,\,p_{2},\,\,p_{3}$) can be
used to obtain four coefficients ($r_{0},\,\,r_{x},\,\,r_{y},\,\,
r_{z}$) that define a single-qubit state. A single-qubit state can
be reconstructed following the steps presented above and an
example is described here. If we obtain $r_{x}=1$,
$r_{y}=\sqrt{3}$, $r_{z}=1$ by the three experimentally measured
probabilities ($p_{1}$, $p_{2}$ and $p_{3}$) on a quantum ensemble
of an unknown charge qubit state $\rho$, then
\begin{eqnarray*}
\rho_{00}&=&\rho_{11}\,=\,\frac{1}{2},\\
\rho_{01}&=&\frac{1}{4}\left(1-i\sqrt{3}\right),\\
\rho_{10}&=&\frac{1}{4}\left(1+i\sqrt{3}\right).
\end{eqnarray*}
Thus, the reconstructed state $\rho$ can be written as
\begin{eqnarray*}
\rho &=&\frac{1}{2}\left(|0\rangle\langle 0|+|1\rangle\langle
1|\right)+\frac{1}{4}\left[\left(1+i\sqrt{3}\right)|1\rangle\langle
0|\right]\nonumber\\
&+&\frac{1}{4}\left[\left(1-i\sqrt{3}\right)|0\rangle\langle
1|\right]
\end{eqnarray*}
whose real $\rho^{\rm (R)}_{ij}$ and imaginary $\rho^{\rm
(I)}_{ij}$ parts are graphically represented in Fig.~\ref{fig3}.

\subsection{Operation time estimates}
The coherent operations required for the tomographic measurements
are limited by the decoherence time $T_{2}$. Now let us explore
whether the single-qubit state can be reconstructed with the
current experiments. To estimate the corresponding time scales for
quantum operations to obtain the measurements of $\sigma_{y}$ and
$\sigma_{x}$, we first take the suggested  parameters from
Ref.~\cite{y}, that is, $E^{0}_{\rm J}=100$ mK (about $8.6
\,\mu$eV or $2.08$ GHz) and $E_{\rm C}=1$ K (about $86 \,\mu$eV or
$20.8$ GHz). Here, we use temperature units for energies as in
reference~\cite{y}. Thus the approximate time scales of one-qubit
operations to obtain $r_{y}$ and $r_{x}$ are
$$t_{x}\approx 5.9\times 10^{-11}\,\,{\rm s}$$ and
$$ t_{y}=t_{z,1}+t_{x,1}+t_{z,2}\approx 7.1\times 10^{-11}\,\,{\rm s}.$$
These time scales, required to reconstruct the single-qubit
states, are within the measured values~\cite{Nakamura,pashkin} of
the decoherence time $T_{2}$ (of the order of magnitude of ns) of
single-qubit charge states.

Now let us consider another set of experimental values. For
example, if the Josephson and charge energies are taken (second
paper in Ref.~\cite{pashkin}) as $2E^{0}_{\rm J}=45 \mu$eV (about
$520$ mK or $10.9$ GHz) and $4E_{\rm C}=580 \mu$eV (about $6.73$ K
or $140$ GHz), then the time scales required to reconstruct
single-qubit states are about $t_{x}\approx2.3\times 10^{-11}$s
and $t_{y}\approx 3.0 \times 10^{-11}$s, which are within the
decoherence time $T_{2}=5$ ns obtained by that
experiment~\cite{pashkin}.

If we take the Josephson and charge energies from in
Ref.~\cite{lehnert}, that is, $2E^{0}_{\rm J}/h=13.0$ GHz (about
$625$ mK or $53.7$ $\mu$eV) and $4 E_{\rm C}/h=149.1$ GHz (about
$7.16$ K or $618$ $\mu$eV), then the time scales required to
reconstruct single-qubit states are about $t_{x}\approx1.9\times
10^{-11}$s and $t_{y}\approx 2.6 \times 10^{-11}$s, which are also
less than one order of magnitude of the decoherence time
$T_{2}=325$ ps measured by that experiment~\cite{lehnert}.

\section{reconstruction of two-qubit states}
\subsection{Theoretical model and two-qubit states}

In this section, we focus on the reconstruction of two-qubit
charge states.  Any two-qubit state $\rho_{1}$ can be
characterized by a density matrix operator
\begin{equation}\label{eq:3}
\rho_{1}=\frac{1}{4}\sum_{i ,j=0,x,y,z}r_{i,j}\,\sigma_{1i}\otimes
\sigma_{2j}
\end{equation}
where the $16$ parameters $r_{i,j}$ are real numbers. The
normalization property of the quantum state requires that
$r_{0,0}=1$, so the state $\rho_{1}$ in Eq.~(\ref{eq:3}) can in
principle be reconstructed~\cite{prove} by $15$ measurements
described by the operators $\sigma_{1i}\otimes\sigma_{2j}$,  where
all $i$ and $j$ are not simultaneously taken to be $0$. If one of
$\sigma_{1i}$ ($i=0, \,x,\,y,\,z$) or $\sigma_{2j}$ ($i=0,
\,x,\,y,\,z$)  is an identity operator among the measurement
operators $\sigma_{1i}\otimes\sigma_{2j}$, we call such a
measurement a single-qubit measurement and only write out the
non-identity Pauli operator in the following expressions. For
example, the operator $\sigma_{1x}\otimes\sigma_{20}$ is called
$\sigma_{x}$ measurement of the first qubit, and abbreviated  by
$\sigma_{1x}$. So there are nine two-qubit measurements among
measurements $\sigma_{1i}\otimes\sigma_{2j}$.  Recall that only
one-qubit is involved during the measurement process  in our
approach. If we want to obtain these nine two-qubit measurements,
then two-qubit operations must be applied~\cite{dpd} such that the
singe-qubit measurement can be equivalently transformed into
expected two-qubit measurements.

Now our task is to find a non-local two-qubit operation and use
this operation to realize all necessary two-qubit measurements on
two-qubit states. Here we consider a model proposed by Makhlin et
al.~\cite{makhlin}, where two charge qubits are coupled  in
parallel to a common inductor with inductance $L$. The
Hamiltonian~\cite{makhlin} is
\begin{eqnarray}\label{eq:4}
H&=&-\,\frac{1}{2}\sum_{l=1}^{2}\left [\delta\!
E_{\rm ch}(n_{l,{\rm g}})\,\sigma_{lz}+{E}_{\rm J}(\Phi_{lx})\,\sigma_{lx}\right]\nonumber\\
&-&E_{\rm
int}(\Phi_{1x},\Phi_{2x})\,\sigma_{1y}\otimes\sigma_{2y},
\end{eqnarray}
where it is assumed that both qubits are identical, so the charge
energies $\delta\! E_{\rm ch}(n_{l,{\rm g}})$ and Josephson
coupling energies $E_{\rm J}(\Phi_{lx})$ take the same form as in
Eq.~(\ref{eq:1}), but now $\delta\!E_{\rm ch}(n_{l,{\rm g}})$ and
$E_{\rm J}(\Phi_{lx})$ for each qubit can be  separately
controlled by the gate voltages and external fluxes. The
interaction energy  $E_{\rm int}$ for two coupled qubits can be
written as
\begin{equation*}
E_{\rm int}(\Phi_{1x},\Phi_{2x})=\frac{E_{\rm J}(\Phi_{1x})E_{\rm
J}(\Phi_{2x})}{E_{\rm L}}
\end{equation*}
with
\begin{equation*}
E_{\rm L}= \left(\frac{C^{0}_{\rm J}}{C_{\rm qb}}\right)^2
\left(\frac{\Phi^{2}_{0}}{\pi^2 L}\right)
\end{equation*}
and $C^{-1}_{\rm qb}=(2C^{0}_{\rm J})^{-1}+C^{-1}_{\rm g}$. Thus,
the interaction between the two qubits can be controlled by two
external fluxes $\Phi_{lx}$ applied to each qubit.

\subsection{Quantum operations and measurements on two-qubit states}
Now, we discuss how to reconstruct two-qubit states from the
experimental measurements $(|1\rangle \langle
1|)_{l}\,\,(l=1,\,2)$. Single charge qubit operations can be
realized by controlling the gate voltage and Josephson couplings.
However the two-qubit operations need to couple a pair of
interacting charge qubits. The realization of the coupling of two
charge qubits have to simultaneously turn on the Josephson
couplings of the two charge qubits in Eq.~(\ref{eq:4}), then
$\sigma_{lx}$ terms have to be included in the two-qubit
operation. However the charge energies for two qubits can be
switched off by applying gate voltages such that $n_{l,{\rm
g}}=1/2$ ($l=1,\,2$), so a two-qubit operation can be governed by
a simpler Hamiltonian
\begin{equation}\label{eq:5}
H^{\prime}=-\frac{1}{2}\sum_{l=1,2} E_{\rm
J}(\Phi_{lx})\,\sigma_{lx} -E_{\rm int}(\Phi_{1x},\Phi_{2x})
\,\sigma_{1y}\otimes\sigma_{2y},
\end{equation}
where  charging energies  are set to zero, $\delta E_{\rm
ch}(n_{l,{\rm g}})=0$ ($l=1,\,2$), with $n_{1,{\rm g}}=n_{2,{\rm
g}}=1/2$, and the external magnetic fields are chosen such that
\begin{equation*}
\Phi_{1x}=\Phi_{2x}\neq \frac{\pi}{2}(2q+1)\Phi_{0}
\end{equation*}
with the positive integer number $q$. The coupling $E_{\rm
int}(\Phi_{1x},\Phi_{2x})$ can be controlled by the external
fluxes $\Phi_{1x}$ and $\Phi_{2x}$.

The basic two-qubit operation can be given by the time-evolution
operator $U(t)=\exp\{-iH^{\prime}t/\hbar\}$, which can be written
by using the Pauli operators as
\begin{eqnarray}\label{eq:6}
&&U(t)\,=\,\frac{1}{2}\left( \cos\phi^{\prime}
+\cos\theta^{\prime} \right) I
+in_{z}\frac{\sin\theta^{\prime}}{2}\left( \sigma _{1x}+\sigma
_{2x}\right) \nonumber\\
&&+\,i\frac{\sin \phi^{\prime} -n_{x}\sin \theta^{\prime}
}{2}\,\sigma_{1z}\otimes \sigma _{2z}\nonumber\\
&&+\,i\frac{\sin \phi^{\prime} +n_{x}\sin \theta^{\prime}}{2}\,
\sigma_{1y}\otimes \sigma _{2y}\nonumber\\
&&\,-\,\frac{\cos\phi^{\prime} -\cos\theta^{\prime} }{2}\,\sigma
_{1x}\otimes \sigma_{2x},
\end{eqnarray}
where
\begin{eqnarray*}
\phi^{\prime}&=&\frac{t}{\hbar}\,E_{\rm int}(\Phi_{1x},\Phi_{2x}),\,\,\,\,\,n_{z}=\frac{a}{\sqrt{1+a^2}},\\
n_{x}&=&\frac{1}{\sqrt{1+a^2}}, \,\,\, \,\,\, \,a=\frac{E_{\rm J}}{E_{\rm int}(\Phi_{1x},\Phi_{2x})},\\
\theta^{\prime}&=&\frac{2}{\hbar}\,E_{\rm
int}(\Phi_{1x},\Phi_{2x})\,\sqrt{1+a^2}\,.\\
\end{eqnarray*}
Since the two external fluxes satisfy the condition
$\Phi_{1x}=\Phi_{2x}$,  we let $E_{\rm J}(\Phi_{1x})=E_{\rm
J}(\Phi_{2x})=E_{\rm J}$ in the expression Eq.~(\ref{eq:6}) for
the two-qubit operation. The physical meaning of the angles
$\theta^{\prime}$ and $\phi^{\prime}$ becomes clearer by virtue of
the ``conjugation-by-$\frac{\pi}{4}\Sigma"$ operation~\cite{lidar}
on the time evolution operator $U(t)$, which is defined as
\begin{equation*}
U^{\prime}(t)=\exp
\left\{i\frac{\pi}{4}(\sigma_{1y}+\sigma_{2y})\right\}U(t)
\exp\left\{-i\frac{\pi}{4}(\sigma_{1y}+\sigma_{2y})\right\},
\end{equation*}
here, $\Sigma=\sigma_{1y}+\sigma_{2y}$. In the conjugate
representation, the time evolution $U^{\prime}(t)$ corresponds to
rotations~\cite{so} around the $y$ axis by an angle
$\phi^{\prime}$ and the $(n_{x},0,n_{z})$ axis by an angle
$\theta^{\prime}$. By choosing the duration $t$ and tuning the
values of $E_{\rm J}$ and $E_{\rm int} (\Phi_{1x}, \Phi_{2x})$, we
can obtain any desired two-qubit operation.

From Eq.~(\ref{eq:3}), it is known that six single-qubit
measurements $\{\sigma_{1i},\sigma_{2j}\}$ with $i,j=x,y,z$ and
nine two-qubit measurements $\{\sigma_{1i}\otimes\sigma_{2j} \}$
with $i,j=x,\,y,\,z$ are enough to obtain fifteen parameters
$r_{i,j}$ of the two-qubit state
$\rho_{1}=\frac{1}{4}\sum_{i,j=0,x,y,z}r_{i,j}\,\sigma_{1i}\otimes
\sigma_{2j}$. The single-qubit measurements $(|1\rangle\langle
1|)_{l}=\frac{1}{2}(\sigma_{l0}-\sigma_{lz})$ ($l=1,\,2$) on a
given state $\rho_{1}$ can be obtained as follows. Two
single-qubit measurements $\sigma_{1z}$ and $\sigma_{2z}$ can be
implemented by the direct measurements $(|1\rangle\langle 1|)_{1}$
and $(|1\rangle\langle 1|)_{2}$ on the given state $\rho_{1}$.
Other four  single-qubit measurements (corresponding to
$\sigma_{1x},\, \sigma_{2x},\, \sigma_{1y}, \,\sigma_{2y}$) need
single-qubit operations.

The single-qubit operations corresponding to measurements
$\sigma_{lx}$ and $\sigma_{ly}$ on two-qubit states are the same
as  measuring $\sigma_{x}$ and $\sigma_{y}$  on single-qubit
states. However, in the single qubit operations, we need to switch
off the interaction of the two qubits.  For example, in order to
obtain the measurement $\sigma_{1y}$, we need to switch off the
interaction between the two-qubit system by setting the applied
external flux $\Phi_{2x}=\pi/2$, and setting the first subsystem
at the degeneracy point and evolving a time $t=
\hbar\pi/4E^{0}_{J}$. Finally,  we make a measurement
$(|1\rangle\langle 1|)_{1}$ on the rotated state, then the
coefficient $r_{y,0}$ can be obtained by this measured result. The
other three measurements can also be obtained by taking
single-qubit operations  similar to $\sigma_{1y}$.

The single-qubit measurements have been obtained by the
measurements $(|1\rangle\langle 1|)_{l} (l=1,2)$ on given states
by using appropriate single-qubit operations as described above.
In order to find out how to obtain the two-qubit measurements via
$(|1\rangle\langle 1|)_{l}$, let us consider the measurements
$(|1\rangle\langle 1|)_{l}$ on the given state $\rho_{1}$
performed by a sequence  $W$ of single-qubit and two-qubit
operations. The corresponding measured probability $p$ can be
expressed as
\begin{equation}
p={\rm Tr}\left[ W\,\rho_{1}\, W^{\dagger}(|1\rangle\langle
1|)_{l}\right]=\frac{1}{2}-\frac{1}{2}{\rm Tr}\left[ \rho_{1}
\,W^{\dagger}\sigma_{lz}W\right],
\end{equation}
where we show that the measurement $(|1\rangle\langle 1|)_{l}$ on
the rotated state $W\rho_{1}\,W^{\dagger}$ may be interpreted as
an equivalent measurement $W^{\dagger}\sigma_{lz}W$ ($l=1,2$) on
the state $\rho_{1}$. So our task now is to find an appropriate
two-qubit operation  and apply this two-qubit and single-qubit
operations to the measured state $\rho_{1}$, such that we can
equivalently obtain the desired two-qubit measurement.

Here, the required two-qubit operation $U(\tau)$ can be obtained
by choosing the evolution time $\tau$, the Josephson coupling
energies $E_{\rm J}$, and $E_{\rm L}$ in Eq.~(\ref{eq:6}) such
that $\phi^{\prime}=(2m-1)\pi/4$ and $\theta^{\prime}=n\pi$ where
$m, \,n$ are positive integers. The above conditions can be
satisfied if the ratio
\begin{equation*}
\frac{E_{\rm L}}{E_{\rm
J}}=\sqrt{\left(\frac{4n}{2m-1}\right)^2-1},
\end{equation*}
and the evolution time $\tau$ is chosen as
\begin{equation*}
\tau=\frac{\hbar\pi}{4E_{\rm J}}\sqrt{(4n)^2-(2m-1)^2}.
\end{equation*}
If we choose the integers $m$ and $n$ to minimize the ratio
$E_{\rm L}/E_{\rm J}$, then $E_{\rm L}/E_{\rm J}=\sqrt{15}\cong
3.87$ when $\theta^{\prime}=\pi$ and $\phi^{\prime}=\pi/4$; so the
two-qubit operation time $\tau$ is chosen as
$\tau=\hbar\pi\sqrt{15}/4E_{\rm J}$. Thus Eq.~(\ref{eq:6}) is
specified by the time evolution operator
\begin{eqnarray}\label{eq:8}
U(\tau)&=&\frac{1}{2\sqrt{2}}\left[\left(1-\sqrt{2}\right)
I-\left(1+\sqrt{2}\right)\,\sigma_{1x}\otimes\sigma_{2x}\right.\nonumber\\
&+&\left.i\,\sigma_{1y}\otimes\sigma_{2y}+i\,\sigma_{1z}\otimes\sigma_{2z}\right].
\end{eqnarray}
Combined with other single qubit rotations, $U(\tau)$ can be used
to obtain all the desired coefficients $r_{i,j}$ corresponding to
the two-qubit measurements $\sigma_{1i}\otimes\sigma_{2j}$, with
$i,\,j=x,\,y,\,z$.

Let us further discuss how to obtain a desired coefficient, for
example, $r_{y,y}$ corresponding to the two-qubit measurement
$\sigma_{1y}\otimes\sigma_{2y}$. We can take the following steps:
\begin{description}
\item (i) \,\, We switch off the interaction between the first and
second qubits by applying an external flux $\Phi_{2x}=\pi/2$,
which means $E_{\rm J}(\Phi_{2x})=0$. Now we only manipulate the
first qubit such that a rotation $\pi/2$ about the $z$ axis,
defined as $Z_{1}=\exp[i\pi\sigma_{1z}/4]$, is performed;  this
single-qubit operation is described in Section II.

\item (ii)\,\, Following the single-qubit rotation $Z_{1}$ of the
first qubit, the gate voltages are applied such that $n_{1,{\rm
g}}=n_{2,{\rm g}}=1/2$, which means that the two qubits work at
the degeneracy points. Simultaneously, we turn on and adjust the
external fluxes so that the external fluxes $\Phi_{lx}$, energies
$E_{\rm L}$ and $E_{\rm J}$ in the two-qubit operation described
by the Hamiltonian (\ref{eq:5}) satisfy the conditions
$\Phi_{1x}=\Phi_{2x}\neq\pi(2q+1)/2$ with positive integer $q$ and
$E_{\rm L}/E_{\rm J}=\sqrt{15}\cong 3.87$. Afterwards, we let the
system evolve a time $\tau=\hbar\pi\sqrt{15}/4E_{\rm J}$; which
means that a two-qubit rotation $U(\tau)$ has been performed.
\end{description}
The operation sequence $W=U(\tau)Z_{1}$ described above changes
state $\rho_{1}$ into
$$\widetilde{\rho}=U(\tau)\,Z_{1}\,\rho_{1}
\,Z^{\dagger}_{1}\,U^{\dagger}(\tau).$$
\begin{description}
 \item (iii)\,\, Finally, when
a single-qubit measurement $(|1\rangle\langle 1|)_{1}$ is
performed on the state $\widetilde{\rho}$, a two-qubit measurement
equivalent to $\sigma_{1z}\otimes\sigma_{2y}$ is implemented:
\end{description}
\begin{equation*}
Z^{\dagger}_{1}\,U^{\dagger}(\tau)\,(|1\rangle\langle
1|)_{1}\,U(\tau)\,Z_{1}=
\frac{1}{2}+\frac{1}{2\sqrt{2}}\left(\sigma_{1z}+\sigma_{1y}\otimes
\sigma_{2y}\right).
\end{equation*}
The corresponding measurement probability $\widetilde{p}$ can be
given as
\begin{eqnarray}
&&\widetilde{p}={\rm Tr}\left\{U(\tau)\,Z_{1}\,\rho_{1}
\,Z^{\dagger}_{1}\,U^{\dagger}(\tau)(|1\rangle\langle 1|)_{1}
\right\}
\nonumber\\
&&=\frac{1}{2}+\frac{1}{2\sqrt{2}}{\rm Tr}\,
[\,\rho_{1}\,(\sigma_{1z}+ \sigma_{1y}\otimes\sigma_{2y})\,]\nonumber\\
&&=\frac{1}{2}+\frac{1}{2\sqrt{2}}(r_{z,0}+r_{y,y}).
\end{eqnarray}
Because the coefficient $r_{z,0}={\rm Tr}
(\rho_{1}\,\sigma_{1z})$, corresponding to the operator
$\sigma_{1z}\otimes \sigma_{20}$, has been given by the
single-qubit measurement $\sigma_{1z}$, then the coefficient
$r_{y,y}={\rm Tr} (\rho_{1}\,\sigma_{1y}\otimes \sigma_{2y})$ is
obtained via $\widetilde{p}$ and  $r_{z,0}$.

In table \ref{tab:table1}, we have summarized nine equivalent
two-qubit measurements described by
$-\sqrt{2}\,W^{\dagger}\sigma_{1z}W$ on the original state
$\rho_{1}$, which are obtained by the first qubit measurement
$(|1\rangle \langle 1|)_{1}$ on the rotated state $W \rho_{1}
W^{\dagger}$ for a sequence $W$ of operations with
appropriately-chosen single-qubit and two-qubit operations. We can
use the results corresponding to these nine equivalent two-qubit
measurements together with the other six single-qubit measurements
to obtain all the coefficients corresponding to the two-qubit
states, and then obtain any two qubit state.

We can also obtain coefficients $r_{ij}$ ($i,\,j\,\neq 0$)
corresponding to all two-qubit measurements by  using the second
qubit measurement $(|1\rangle\langle 1|)_{2}$.  For example, if we
make a measurement $(|1\rangle\langle 1|)_{2}$ on the rotated
state $\widetilde{\rho}$ considered above, we obtain another
equivalent two-qubit measurement, which is expressed as
\begin{equation*}
Z^{\dagger}_{1}\,U^{\dagger}(\tau)\,(|1\rangle\langle
1|)_{2}\,U(\tau)\,Z_{1}=\frac{1}{2}+\frac{1}{2\sqrt{2}}(\sigma_{2z}-\sigma_{1x}\otimes\sigma_{2x}).
\end{equation*}
Using this measurement, combined with the single-qubit measurement
$\sigma_{2z}$, we can obtain the coefficient $r_{x,x}$
corresponding to the two-qubit measurement
$\sigma_{1x}\otimes\sigma_{2x}$.  Nine equivalent two-qubit
measurements realized by  the second qubit $(|1\rangle\langle
1|)_{2}$ have also been summarized in Table \ref{tab:table2}.
Comparing tables \ref{tab:table1} and \ref{tab:table2} shows that
different operations and steps are required in order to obtain the
same coefficient for different measurements. For example, in order
to obtain  $r_{x,z}$, two operation steps are needed for the first
qubit measurement $(|1\rangle\langle 1|)_{1}$, but it needs four
steps for the second qubit measurement $(|1\rangle\langle
1|)_{2}$.
\begin{figure}
\includegraphics[width=7.8cm]{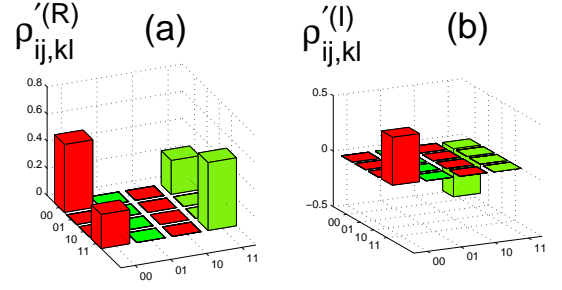}
\caption[1]{Graphical representation of the density matrix
$\rho^{\prime}$ for the two-qubit state described in the example
given in section III. The real $\rho^{\prime \rm (R)}_{ij,kl}$ and
imaginary $\rho^{\prime \rm (I)}_{ij,kl}$ parts of the density
matrix elements for the two-qubit state $\rho^{\prime}$ in the
basis $|00\rangle, \, |01\rangle, \, |10\rangle, \, |11\rangle$
are plotted in (a) and (b) respectively.}\label{fig4}
\end{figure}
\subsection{An example}
We can also give another schematic example for a reconstructed
two-qubit state. For instance, according to the operations steps
discussed above for the reconstruction of any two-qubit state, if
we obtain $r_{x,x}=1/8$, $r_{x,y}=r_{y,x}=\sqrt{3}/8$,
$r_{z,z}=1/4$ and $r_{y,y}=-1/8$ from the sixteen measured
probabilities on an ensemble of identically prepared copies of a
two-qubit system with unknown state $\rho^{\prime}$, then we can
reconstruct this unknown state as
\begin{eqnarray*}
\rho^{\prime}&=&\frac{1}{2}[|00\rangle \langle 00|+|11\rangle
\langle
11|]+\frac{1}{2}\left(1-i\sqrt{3}\right)|00\rangle\langle 11|\nonumber\\
&+&\frac{1}{2}\left(1+i\sqrt{3}\right)|11\rangle\langle 00|,
\end{eqnarray*}
which is graphically shown in Fig.\!~(\ref{fig4}) with the real
$\rho^{\prime \rm (R)}_{ij,kl}$ and imaginary $\rho^{\prime \rm
(I)}_{ij,kl}$ parts of the reconstructed state $\rho^{\prime}$,
where $i,\,j,\,k,\,l$ can take the values  $0$ or $1$.

\subsection{Operation time estimates}
We can also estimate the operation time required to reconstruct
two-qubit states for the Josephson and charge energies~\cite{y}
$E^{0}_{\rm J}=100$ mK and $E_{\rm C}=1$ K. We assume that the
ratio $E_{\rm L}/E_{\rm J}=\sqrt{15}\cong3.87$ is obtained by
adjusting the external flux $\Phi_{lx}$ ($l=1,\,2$) such that
$\Phi_{lx}=0$, which means the ratio between $E_{\rm L}$ and
$E^{0}_{J}$ should satisfy the condition $E_{\rm L}/E^{0}_{\rm
J}=2\sqrt{15}\cong7.74$ when the circuits are fabricated. In such
case, the realization of the two-qubit operation in
Eq.~(\ref{eq:8}) requires a time $\tau\approx 2.32\times 10^{-10}$
s. Our previous estimates for the times to perform $\pi/2$
rotations about the $x$ and $z$ axes are $5.9\times 10^{-11}$ s
and $3.0\times 10^{-12}$ s, respectively. Then, using tables
\ref{tab:table1} and \ref{tab:table2},  we can estimate the total
operation time required for obtaining the coefficients of the
two-qubit measurements corresponding to the first or second qubit
measurements, respectively. We find that the required operation
times for the two-qubit measurements are less than $0.4$ ns for
the two-qubit measurements. The decoherence time $T_{2}$ (e.g.,
the decoherence time of charge qubit is about $5$ ns in
reference~\cite{Nakamura}) experimentally obtained shows that it
is possible to reconstruct two-qubit states within the current
measurement technology.

At present, completely controllable multi-qubit superconducting
circuits are not experimentally achievable. Here, let us consider
the operation time estimates based on another controllable
model~\cite{you}. In this model, $N$ charge qubits are coupled to
a common superconducting inductance $L$. The Hamiltonian of any
pair of qubits, say $i$ and $j$, is
\begin{equation}
H^{\prime}=\sum_{k=i,j}(\varepsilon_{k}\sigma^{(k)}_{z}+\epsilon_{k}\sigma^{(k)}_{x})
+\chi \sigma^{(i)}_{x}\otimes \sigma^{(j)}_{x},
\end{equation}
where the coupling constant $\chi$ can be tuned to zero by
changing  the flux either through the common inductance $L$, or
through the qubit $i$ (or $j$). Moreover, the parameters
 $\varepsilon_{k}$ and  $\epsilon_{k}$ are respectively controlled
by the voltage applied to the $k$th qubit and the magnetic flux
through the $k$th qubit. The conditional logic gates, e.g.,
controlled-NOT and controlled-phase-shift gates, can be performed
by virtue of only one two-bit operation and also single-qubit
operations in this circuit. This approach is more accessible to
experiments, facilitating tomographic measurements. According to
calculations~\cite{liu} of tomographic measurements for a class of
representative quantum computing models of  solid state systems,
the two-qubit operation required for the realization of the
multi-qubit measurements in this circuit can be easily obtained.
That is, if the ratio between the Josephson energy $E^{0}_{\rm J}$
and the two-qubit coupling energy $\chi$ is $E^{0}_{\rm J}=2\chi$,
when the circuit is fabricated, then a two-qubit operation
$\widetilde{U}(\tau^{\prime})=-i\sigma_{1x}\otimes \sigma_{2x}$
can be obtained  with the evolution time $\tau^{\prime}\approx
1.2\times 10^{-10}$ s when the Josephson energy is taken as
$E^{0}_{\rm J}=100$ mK. Here, we assume that the two charge qubits
are identical and the Josephson energies are maximum when the
two-qubit operation is performed If the charging energy is taken
as $E_{\rm C}=1$ K, then $\pi/2$ rotations around the $z$ and $x$
axes need times $3.0\times 10^{-12}$ s and  $5.9\times 10^{-11}$
s, respectively. The operations to get each of the sixteen
(single- and two-qubit) measurements can also be obtained for this
model by using an approach similar to the one described above, the
estimated operation times to obtain all coefficients of the
two-qubit measurements are less than $0.3$ ns, which is also
within the experimentally obtained decoherence time $T_{2}=5$ ns.

\begin{table}
\caption{\label{tab:table1}Equivalent two-qubit measurements
$-\sqrt{2}\,W^{\dagger}\,\sigma_{1z}\, W$ obtained by measuring
$(|1\rangle\langle 1|)_{1}$ on the state $W\rho_{1}\, W^{\dagger}$
with a sequence of appropriately-chosen quantum operations $W$.}
\begin{ruledtabular}
\begin{tabular}{ccc}
Two-qubit & Quantum & Equivalent two-qubit\\
measurement &operation\footnote {$X_{l}$ and $Z_{l}$ denote single
qubit rotations $\pi/2$ of $l$th  qubit  about the $x$ and $z$
axes, respectively, and
$\tau=\hbar\pi\sqrt{15}/4E_{\rm J}$.}\,\, $W$&  measurement\\
\hline
 $\sigma_{1x}\otimes\sigma_{2y}$& $U(\tau)$&
$\sigma_{1z}+\sigma_{1x}\otimes\sigma_{2y}$\\
  \hline
$\sigma_{1x}\otimes\sigma_{2z}$&$X_{1}U(\tau)$&$-\sigma_{1y}+
 \sigma_{1x}\otimes\sigma_{2z}$\\
\hline
$\sigma_{1x}\otimes\sigma_{2x}$&$U(\tau)Z_{2}$&$\sigma_{1z}-\sigma_{1x}\otimes\sigma_{2x}$\\
\hline
$\sigma_{1y}\otimes\sigma_{2y}$&$U(\tau)Z_{1}$&$\sigma_{1z}+\sigma_{1y}\otimes\sigma_{2y}$\\
\hline
$\sigma_{1y}\otimes\sigma_{2z}$&$X_{1}U(\tau)Z_{1}$&$\sigma_{1x}+\sigma_{1y}\otimes\sigma_{2z}$\\
\hline
$\sigma_{1y}\otimes\sigma_{2x}$&$U(\tau)Z_{1}Z_{2}$& $\sigma_{1z}-\sigma_{1y}\otimes\sigma_{2x}$\\
\hline $\sigma_{1z}\otimes\sigma_{2y}$&$U(\tau)Z_{1}X_{1}$&$-\sigma_{1y}+\sigma_{1z}\otimes\sigma_{2y}$\\
\hline $\sigma_{1z}\otimes\sigma_{2z}$&$X_{1}U(\tau)Z_{1}X_{1}$&$\sigma_{1x}+\sigma_{1z}\otimes\sigma_{2z}$\\
\hline $\sigma_{1z}\otimes\sigma_{2x}$&$U(\tau)Z_{1}Z_{2}X_{1}$&$-\sigma_{1y}-\sigma_{1z}\otimes\sigma_{2x}$\\
\end{tabular}
\end{ruledtabular}
\end{table}

\begin{table}
\caption{\label{tab:table2}Equivalent two-qubit measurements
$-\sqrt{2}\,W^{\dagger}\,\sigma_{2z} W$ obtained by measuring
$(|1\rangle\langle 1|)_{2}$ on the state
$W\,\rho_{1}\,W^{\dagger}$ with a sequence of appropriately-chosen
quantum operations $W$.}
\begin{ruledtabular}
\begin{tabular}{ccc}
Two-qubit & Quantum & Equivalent quantum\\
measurement &operation $W$&  measurement\\
\hline
$\sigma_{1x}\otimes\sigma_{2x}$&$U(\tau)Z_{1}$&
$\sigma_{2z}-\sigma_{1x}\otimes\sigma_{2x}$\\
\hline
$\sigma_{1y}\otimes\sigma_{2x}$&$U(\tau)$&
$\sigma_{2z}+\sigma_{1y}\otimes\sigma_{2x}$\\
 \hline
$\sigma_{1z}\otimes\sigma_{2x}$& $U(\tau)X_{1}$&
$-\sigma_{2y}+\sigma_{1z}\otimes\sigma_{2x}$\\
\hline
 $\sigma_{1x}\otimes\sigma_{2y}$&$U(\tau)Z_{1}Z_{2}$&$\sigma_{2z}-\sigma_{1x}\otimes\sigma_{2y}$\\
 \hline
$\sigma_{1y}\otimes\sigma_{2y}$
&$U(\tau)Z_{2}$&$\sigma_{2z}+\sigma_{1y}\otimes\sigma_{2y}$\\
\hline
$\sigma_{1z}\otimes\sigma_{2y}$&$U(\tau)X_{1}Z_{2}$&$\sigma_{2x}+\sigma_{1z}
\otimes\sigma_{2y}$\\
 \hline
$\sigma_{1x}\otimes\sigma_{2z}$&$U(\tau)Z_{1}Z_{2}X_{2}$&$-\sigma_{2y}-\sigma_{1x}\otimes\sigma_{2z}$\\
\hline
$\sigma_{1y}\otimes\sigma_{2z}$
&$U(\tau)Z_{2}X_{2}$& $-\sigma_{2y}+\sigma_{1y}\otimes\sigma_{2z}$\\
\hline
$\sigma_{1z}\otimes\sigma_{2z}$&$U(\tau)X_{1}Z_{2}X_{2}$&$\sigma_{2x}+\sigma_{1z}
\otimes\sigma_{2z}$
\end{tabular}
\end{ruledtabular}
\end{table}

\section{ Reconstruction of multiple qubit states}

In the above two sections, we focused on the reconstruction of the
single and two qubits states. In this section, we discuss the
reconstruction of any $n$-qubit state. In the multiple qubit
charge circuit, the dynamical evolution is governed by the
Hamiltonian~\cite{makhlin}
\begin{eqnarray}\label{eq:10}
H&=&-\,\frac{1}{2}\,\sum_{l=1}^{n} [\delta\!
E_{\rm ch}(n_{l,{\rm g}})\,\sigma_{lz}+{E}_{\rm J}(\Phi_{lx})\,\sigma_{lx}]\nonumber\\
&-&\sum_{l<k}E_{\rm
int}(\Phi_{lx},\Phi_{kx})\,\sigma_{ly}\otimes\sigma_{ky},
\end{eqnarray}
where $\delta\!E_{\rm ch}(n_{l,{\rm g}})=4E_{\rm C}(1-2n_{l,{\rm
g}})$, $E_{\rm J}(\Phi_{lx})=2E^{0}_{\rm
J}\cos(\pi\Phi_{lx}/\Phi_{0})$, and $E_{\rm
int}(\Phi_{lx},\Phi_{kx})$ take  the same form as in
Eq.~(\ref{eq:4}). We also assume $E_{\rm L}/2E^{0}_{\rm
J}=\sqrt{15}\cong3.87$ and the single-qubits  are nominally
identical. By virtue of the controllable
Hamiltonian~(\ref{eq:10}),  in principle we can use $(n-1)$
two-qubit operations together with some single-qubit operations to
reconstruct any $n$-qubit state, which can also be described by
the density matrix operator
\begin{equation*}
\rho_{2}=\frac{1}{2^{n}}\sum_{l_{1},l_{2},\cdots,
l_n=0,x,y,z}r_{l_{1},l_{2},\cdots,l_{n}}\,\sigma_{l_{1}}\otimes\sigma_{l_{2}}\cdots\otimes
\sigma_{l_{n}} \end{equation*} with $2^{n}$ real parameters
$r_{l_{1},l_{2},\cdots,l_{n}}$ corresponding to the measurements
$\sigma_{l_{1}}\otimes\sigma_{l_{2}}\cdots\otimes \sigma_{l_{n}}$.
But, here, we only show how to obtain a coefficient corresponding
to a three-qubit measurement. The generalization to obtain
coefficients of multiple qubit measurements is straightforward.

In order to determine a three-qubit state, we need to make,
single-qubit, two-qubit, and three-qubit measurements. It is known
that all coefficients corresponding to single-qubit and two-qubit
measurements can be obtained by using the same operations and
measurements $(|1\rangle\langle 1|)_{l=1,\,2\,3}$ as in  section I
and II. When we make two-qubit operations on, for example, the
first and second qubits,  the interaction of the third qubit with
these two qubits is switched off by the applied flux
$\Phi_{3x}=\pi/2$. Now let us show how to obtain the coefficients
corresponding to the three-qubit measurements. For example, for
the coefficient $r_{x,z,y}$ of the measurement
$\sigma_{1x}\otimes\sigma_{2z}\otimes\sigma_{3y}$, we should make
the following sequence of quantum operations:
\begin{description}
\item (i) \,\,Switch off the interaction of the third qubit  with
the first and second qubits by applying the  flux
$\Phi_{3x}=\pi/2$. Then make a two-qubit operation $U_{12}(\tau)$,
with the same form as Eq.~(\ref{eq:8}). We use the subscript ``12"
to denote two-qubit operations on the first and second qubits.

\item (ii) \,\,  Switch off the interaction between the first and
second qubits by setting $\Phi_{2x}=\pi/2$, and making a $\pi/2$
rotation about the $z$ axis for the first qubit.

\item  (iii)\,\, Make another two-qubit rotation $U_{13}(\tau)$ on
the first and third qubits by adjusting the external fluxes such
that $\Phi_{1x}=\Phi_{3x}=0$. The two-qubit operation
$U_{13}(\tau)$ takes the same form as Eq.~(\ref{eq:8}), but the
subscript ``$2$" of the Pauli operators in Eq.~(\ref{eq:8}) is
replaced by the subscript ``$3$". This process can be described as
\begin{eqnarray}
\rho_{2}&&\,\xrightarrow{U_{12}(\tau)}\,U_{12}(\tau)\, \rho_{2}
\,U^{\dagger}_{12}(\tau)\,\xrightarrow{Z_{1}}\,Z_{1}\,U_{12}(\tau)
\,\rho_{2}\,
U^{\dagger}_{12}(\tau)\,Z^{\dagger}_{1}\nonumber\\
&&\,\xrightarrow{U_{13}}\,U_{13}\,Z_{1}\,U_{12}(\tau)\,\rho_{2}\,
U^{\dagger}_{12}(\tau)\,Z^{\dagger}_{1}\,U^{\dagger}_{13}\,.
\end{eqnarray}

\item (iv)\,\, Finally, make a measurement  $(|1\rangle\langle
1|)_{1}$ on the above rotated state,  and obtain the equivalent
measurement
\begin{eqnarray}
&&U^{\dagger}_{12}(\tau)\,Z^{\dagger}_{1}\,U^{\dagger}_{13}\,(|1\rangle\langle
1|)_{1}\,U_{13}\,Z_{1}\,U_{12}\,=\frac{1}{2}-\frac{1}{4}\sigma_{1z}+\nonumber\\
&&+\frac{1}{4}(\sigma_{1x}\otimes\sigma_{2y}+\sigma_{1y}\otimes\sigma_{3y}-\sigma_{1x}\otimes
\sigma_{2z}\otimes\sigma_{3y}),
\end{eqnarray}
\end{description}
and corresponding measurement result $p^{\prime\prime}$ is
\begin{equation}
p^{\prime\prime}\,=\,\frac{1}{2}-\frac{r_{z,0,0}+r_{x,y,0}+r_{y,0,y}-r_{x,z,y}}{4}\,.
\end{equation}
Finally, we can obtain the coefficient $r_{x,y,z}$  based on
$p^{\prime\prime}$ and the single  and two qubit measurement
results $r_{z,0,0}$, $r_{x,y,0}$ and $r_{y,0,y}$, which can be
obtained by using the same way described in sections II and III.
Other coefficients corresponding to three-qubit measurements can
also be obtained by using a similar procedure. According to the
estimated time for reconstructing the two-qubit states, we believe
that it is also possible to reconstruct the three-qubit states
using  current technology. Any multiple-qubit can also be
reconstructed by sequentially designing the single-qubit and
two-qubit operations. The generalization to multiple-qubit is an
extension of the procedure that we outlined above.

\section{quantum process tomography}

It is worth briefly reviewing that, based on qubit {\it state}
tomography, the noisy channel (usually denoted as the ``black
box") of the controllable charge qubits can also be determined.
This experimental determination of the dynamics of the ``black
box" is called  quantum {\it process} tomography~\cite{il}, which
can be described as follows:
\begin{description}
\item (i) \,\,
 Many known quantum states of the
system under investigation are input into the ``black box", which
is an unknown quantum channel, for example, an arbitrary
environment. \item (ii) \,\,After a certain time, the output
states evolve into unknown states. \item (iii) \,\,By using the
state tomography, we can ascertain these unknown states. \item
(vi) \,\,Finally, an unknown quantum channel is determined by the
data obtained  for the tomographic measurements on these states.
\end{description}
Experimentally, in order to determine the noisy channel of the
studied $N$-qubits~\cite{il}, $N^2$ known states need to be
prepared, and these states must have density matrices which span
the space of any allowed input state density matrices.

We have shown that  single-qubit {\it state} tomography is
experimentally accessible. In order to perform quantum {\it
process} tomography for a single charge qubit. Four kinds of
different charge states  $|0\rangle$, $|1\rangle$,
$(|0\rangle+|1\rangle)/\sqrt{2}$, and
$(|0\rangle+i|1\rangle)/\sqrt{2}$ need to be experimentally
prepared. These states can be generated in a SQUID-based charge
qubit with current experiments~\cite{Nakamura,pashkin,lehnert}.
Thus, the {\it process} tomography of a single charge qubit is
achievable using current technology. With further developments of
this technique, the {\it process} tomography of multiple charge
qubits could also be realized, when data from multi-qubit state
tomography is obtainable.

\section{conclusions}

In conclusion, we discuss how to reconstruct charge qubit states
via one-qubit measurements using controllable superconducting
quantum devices. Detailed operations for reconstructing single-
and two-qubit states are presented. Any $n$-qubit state can also
be reconstructed by using $n-1$ two-qubit operations similar to
Eq.~(\ref{eq:8}) for different qubit pairs and combining these
with required single-qubit operations. Thus the non-local
two-qubit operation Eq.~(\ref{eq:8}) plays a key role in the
reconstruction of the multiple-qubit states. However, this
two-qubit operation is not unique for achieving our purpose. We
should note that operations to obtain a fixed coefficient
corresponding to multiple-qubit measurements are not unique. The
measurements $(|1\rangle\langle 1|)_{l}$ ($l=1,\,2,\,\ldots, \,n$)
on the given state with fixed operations $W$ are different for
each qubit $l$, because $W$ is not symmetric when exchanging $l$.
Our proposal can also be generalized to other superconducting
charge qubit circuits with the coupling mediated by photons or a
tunable oscillator, e.g., Refs.~\cite{you2,wei}, or other types of
superconducting qubits, e.g.,
Refs.~\cite{majer,izmalkov,xu,berkley,mcdermott}.

We find that the longest operation times  to obtain the
coefficient of single-qubit and two-qubit states are of the order
of $0.01$ ns and  $0.4$ ns, respectively,  which is less than the
decoherence time~\cite{Nakamura} $T_{2}=5$ ns.  Moreover, the
$\pi$ and $\pi/2$ pulses for single-qubit operations can be
performed very well, e.g., in the experiments of the charge echo
~\cite{echo}, and NMR-like experiments~\cite{esteve}. Another
experimental estimate shows us that the manipulation accuracy can
reach $80-90 \%$ (e.g., as in the second reference of
Ref.~\cite{Nakamura}).  Thus the single-qubit states could be
reconstructed and the process tomography should also be accessible
in single-qubit charge systems with current experimental
capabilities. In principle, the two-qubit states can also be
reconstructed by virtue of well-controlled time for the two-qubit
operation. We should also note that larger values of the charge
energy $E_{\rm ch}$, the Josephson energy $E^{0}_{\rm J}$, and
coupling energy $E_{\rm int}(\Phi_{lx},\Phi_{kx})$ can make the
operation times shorter. Thus these larger values should be
realized in order to facilitate the tomographic reconstruction.

Quantum oscillations and conditional gate operations have been
demonstrated in two coupled charge qubits with the
interactions~\cite{pashkin} always turned on. Completely
controllable two-qubit charge systems have not been realized yet.
However, the coupled two charge qubits, allowing on and off
switching of the interaction, might be realizable in the
future~\cite{averin}.  Then our proposal will become realizable.
Because the unswitchable two-qubit interaction makes single-qubit
operations impossible, our proposed scheme cannot be readily used
to the experimental reconstruction of multiple-qubit charge states
when the two-qubit interactions are always turned on. However, for
the two-qubit circuit with ``always-on" interaction, most of the
single-qubit parameters~\cite{majer,izmalkov,xu,berkley,mcdermott}
can be tuned.  We can adjust these parameters to obtain $15$
different two-qubit operations, and then derive $15$ different
measurement equations with these operations on input states.
Afterwards, the two-qubit states can finally be determined. The
details on how to reconstruct the superconducting two-qubit states
with the ``always-on" couplings  will be presented elsewhere.
However, how to reconstruct qubit states  in  multiple-qubit (more
than two qubits) circuits with ``always-on" interactions is an
open problem.

\section{discussions}
In our paper, to simplify the algebra, we focus on one particular
family of measurements which are constructed by the direct product
of the Pauli operators. However, one can conceive that other
complete sets of measurements can also be used to do tomography.
These different complete sets of measurements can be transformed
to each other by unitary operators. In practice, within the
duration of the controllable manipulation, the smaller Bloch
rotation might be advantageous to speed up the measurements, but
it might also decrease the accuracies of the measurements due to a
longer measuring time. How to choose suitable sets of operations
during the measurement process is an important technical question
for the measurement.

It should also be pointed out that here we discuss an ideal case.
In practice, the environmental effect is unavoidable, which result
in the relaxation (characterized by $T_{1}$) and decoherence
(characterized by $T_{2}$) of the qubits. For example, in the
single-qubit state tomography, non-negligible $T_{1}$ decreases
all three probabilities of the measurements, however
non-negligible $T_{2}$ reduces the probabilities of the
measurements with rotations about $x$ and $y$
directions~\cite{book}. So the environmental effect on the
reconstructed states is required to be considered in practice for
more specific model. Further, the required quantum operations,
especially two-qubit nonlocal operation, are difficult to
accurately implement during the process of experiments. For
example, the probabilities of theoretical calculations with qubit
operations for the first and second qubit measurements
$(|1\rangle\langle 1|)_{1}$ and $(|1\rangle\langle 1|)_{2}$ are
related to parameters $r_{i,j}$ of the equivalent measurements
shown in tables~\ref{tab:table1} and ~\ref{tab:table2}, however,
the measuring results of inaccurately experimental two-qubit
operations will actually relate to not only these results shown in
tables~\ref{tab:table1} and ~\ref{tab:table2}, but also other
extra terms. If these extra terms are not negligible, the
reconstructed states might violate the properties of the positive
semi-definiteness of the physical state $\rho$. A third error
source is the imperfect readout of the charge qubit (for
experiments, e.g., using single charge qubit~\cite{echo}, the
fidelity of the readout can reach $99 \%$). The limited
statistical data also affect the reconstruction of the states. All
these imperfection can make the reconstructed states violate the
important basic properties of the physical states: normalization,
Hermiticity, and positivity. In order to reconstruct a physical
qubit state, in principle the maximum likelihood estimation of
density matrices can be employed to minimize experimental errors.
This method can be applied to numerically optimize the
experimental data, which has been used in the optical
systems~\cite{mj} and more detailed discussions on this method can
be referred to a very good Ref.~\cite{book2}.

When the tomography is processed, the external flux applied to the
SQUID needs to be very quickly changed. For instance, the duration
for changing $\Phi_{0}/2$ within a SQUID loop should at least be
less than the decoherence time. Thus a pulse field magnetometer
with a rapid sweep rate may be required in this experiment. If the
sweep rate~\cite{A} of the pulse field magnetometer reaches, e.g.
$10^{8}$ Oe/s, then the time to change $\Phi_{0}/2$ in the loop
needs about $0.25$ ns for a SQUID area of $400 (\mu{\rm m})^2$.

We also notice that the number of rotations for the measured
density matrix elements to a preferable direction (e.g. $y$
instead of $z$) grows exponentially with the number of qubits. How
to solve this problem is still an open question.

\section{acknowledgments}

We thank X. Hu, J.Q. You, Y.A. Pashkin, O. Astafiev,  and J.S.
Tsai for their helpful comments and discussions. This work was
supported in part by the National Security Agency (NSA) and
Advanced Research and Development Activity (ARDA) under Air Force
Office of Research (AFOSR) contract number F49620-02-1-0334, and
by the National Science Foundation grant No. EIA-0130383.


\begin{thebibliography}{99}
\bibitem{comments}Recently,  simultaneous measurements with two qubits
were performed in a phase qubit circuit~\cite{mcdermott}.
\bibitem{prove}If we make a measurement $\sigma_{1i}\otimes\sigma_{2j}$
 on multi-qubit state $\rho$, then the measurement probability $p={\rm
Tr}(\rho\sigma_{1i}\otimes\sigma_{2j})=\frac{1}{4}\sum_{i,j=0}^{3}r_{i,j}{\rm
Tr}(\sigma_{1i^{\prime}}\sigma_{1i}){\rm
Tr}(\sigma_{2j^{\prime}}\sigma_{2j})=r_{i,j}\delta_{i
i^{\prime}}\delta_{j j^{\prime}}.$
\end{thebibliography}
\end{document}